\documentclass[twocolumn,showpacs,preprintnumbers,amsmath,amssymb]{revtex4}
\usepackage{graphicx}
\usepackage[]{bm}
\def\vec#1{{\bm{#1}}}
\def\e{{\rm e}}

\def\e#1{\,{\rm e}#1}
\def\ket#1{|#1\rangle }
\def\bra#1{\langle #1|}
\def\braket#1#2{\langle #1|#2\rangle}
\def\i{{\bf i}}
\def\Tr{\mathop{{\rm Tr}}}
\begin{document}
\draft

\title{ $U(1)$ symmetry breaking in one-dimensional Mott insulator
  studied by the Density Matrix Renormalization Group method}

\author{Isao Maruyama, Tetsuji Koide and Yasuhiro Hatsugai}
\address{
Department of Applied Physics, University of Tokyo, Hongo Bunkyo-ku, Tokyo 113-8656, Japan
}
\begin{abstract}
  A new type of external fields violating the particle number
  preservation is studied in one-dimensional strongly
  correlated systems by the Density Matrix Renormalization Group
  method.  Due to the U(1) symmetry breaking, the ground state has
  fluctuation of the total particle number, which implies injection of
  electrons and holes from out of the chain.  This charge fluctuation can
  be relevant even at half-filling because the particle-hole symmetry
  is preserved with the finite effective field.  In addition, we 
  discuss a quantum phase transition obtained by considering the
  symmetry-breaking fields as a mean field of interchain-hopping.
\end{abstract}
\maketitle

\section{Introduction}
Doped Mott insulators is one of possible candidates of superconductors
with electron-electron correlation. At a rational filling, strong
electron-electron interaction makes electrons localized in real space.
This is the Mott insulator where charge excitations are gapped.  Even
at this rational filling, the spin degree of freedom survives as a
gapless mode where quantum objects as collection of the $S=1/2$ spins
form a singlet ground state.

When mobile carriers are introduced into the Mott insulator, we may
expect that the charge gap is destroyed which realizes the
superconducting ground state driven by the electron interaction.  In
the resonating valence bond (RVB) picture proposed by Anderson to
describe high-$T_c$ superconducting cuplates\cite{Science.235.1196,
  PRB.37.3664} , doped holes itinerate in a spin-singlet ground state
and condense into a superconducting state.  If there is no doped hole,
i.e., half-filled case, the spin-singlet ground state is expected as a
Mott insulator.

Apart from real doping, that is, changing chemical potential, there
can be several possibilities for the effective carrier doping.  One of
a natural possibility can be geometrical frustration in layered
organic superconductors, which is approximated by a half-filled Hubbard
model with next-nearest-neighbor hopping\cite{cond-mat.0607470}.
Another one is gossamer superconductivity by Laughlin
\cite{cond-mat.0209269}.  Even at the half-filed case, finite double
occupancy may destroy the Mott insulator at a small on-site Coulomb
repulsion and lead the ground state to the gossamer superconducting
state\cite{PRL.90.207002, PRB.71.014508}.  This theory has been also
applied to organic superconductors\cite{PRL.94.067005}.  The gossamer
superconducting state is realized not as a ground state of strongly
correlated system but a variational function based on a Bardeen Cooper
Schrieffer (BCS) superconducting state.  However, it can be an exact
ground state of the model Hamiltonian \cite{cond-mat.0209269} which
violates the charge conservation as the BCS Hamiltonian.  It is thus
theoretically interesting to consider $U (1)$ symmetry breaking
generically.

Let us recall the BCS Hamiltonian.  The BCS Hamiltonian with quadratic
terms $\Delta_{k} c_{k\uparrow}^\dagger c_{-k\downarrow}^\dagger$ has
been widely accepted as a theoretical model for superconductors, where
$\Delta_{k}$ is a mean field of pair annihilation amplitude $\langle
c_{k\uparrow} c_{-k\downarrow}\rangle$ and $c_{k\sigma}$ are
annihilation operators of Fermion.  The mean field violates the $U
(1)$ gauge symmetry, i.e., the total particle number is not preserved
but fluctuating.  This charge fluctuation turns out to diverge in the
thermodynamic limit\cite{M.Tinkham}.

The most simple candidate of $U (1)$ symmetry breaking terms is
$\Delta_{k} c_{k}^\dagger + h.c.$, which implies injection of
electrons and holes from out of the system.  The concept of this term
is directly connected to local charge fluctuation or doping.  Although
the previous study is limited to free fermion, such a 1D system
defined as $H= \sum_k k c_k^\dagger c_k + \Delta \sum_k (c_k^\dagger +
c_k)$ has been studied and was solved with a Jordan-Wigner
transformation\cite{PRB.32.7518} and with a canonical
transformation\cite{PRB.51.1743}.  To clarify the meaning of $\Delta$,
let us describe the procedure of Ref.\cite{PRB.32.7518} in detail;
this spinless fermion chain can be mapped to semi-infinite $XY$ model
with a local magnetic field at the boundary.  This local magnetic
field in the $xy$ plane turns out to be $\Delta$.  In
Refs. \cite{PRB.32.7518, PRB.51.1743}, $\Delta$ terms were introduced
in different contexts.  It is a common concept that the Hamiltonian is
an effective one after tracing out of some environment.

There are two motivations of the present work.  The first is to
clarify the properties of strongly correlated systems with nonuniform
charge fluctuation.  Especially, a particle-hole symmetric Hubbard
model with $\Delta$ will make it possible to realize the ground state
which is a linear combination of half-filled Mott insulating state
with electron-doped and hole-doped states, which evokes the gossamer
theory.  It is interesting to evaluate fluctuation of total particle
number as a direct effect of the $U (1)$ symmetry breaking.  The other
is to test a mean-field type approach for the interchain hopping of
fermion chains.  When we consider a decoupling of interchain hopping
$t_\perp c_{i\sigma}^\dagger c_{\perp i\sigma}$ into $\Delta_{i\sigma}
c_{i\sigma}^\dagger$ with the mean-field type approximation, the
effective fields $\Delta_{i\sigma}$ can be identified as $\langle
t_\perp c_{i\sigma} \rangle$.\footnote{ The limitation of this
  approach, for example the replacement of a fermionic operator by
  c-number, is discussed in \S \ref{sec:conc}.  } In this approach,
the Hamiltonian is considered as an effective one obtained after
tracing out neighboring chains in the quasi-one-dimensional (quasi-1D)
systems.

Since the Hubbard model has Coulomb interaction, we adopt the Density Matrix
Renormalized Group (DMRG) method, which is one of powerful numerical
methods for 1D quantum systems\cite{RMP.77.259, JPA.36.381,
  cond-mat.0404212}.  The DMRG method has achieved high precision in
various 1D systems while application to two-dimensional lattice is
considered to be difficult.  As an application to higher dimensions,
we note that the DMRG method has already applied to quasi-1D spin
systems with interchain couplings as mean fields in the manner of
Ref.\cite{PRB.11.2042}.  This paper may be first attempt to quasi-1D
fermionic systems with interchain hopping as mean fields.

The paper is organized as follows.  In \S~\ref{sec:model} we construct
a Hamiltonian with a generalized $U (1)$ breaking term and mention a
``bath'' site introduced by the canonical
transformation\cite{PRB.51.1743}.  In \S \ref{sec:method}, we describe
an application of the DMRG method to the Hamiltonian which does not
conserve particle number.  In \S \ref{sec:results}, we demonstrate the
mean-field type approach for quasi-1D strongly correlated electron
systems.  Finally we conclude the results and discuss problems of the
mean-field type approach.  In Appendix, the note for the canonical
transformation is summarized.

\section{Correlated Electron systems with $U(1)$ symmetry breaking term\label{sec:model}}
Let us define a Hamiltonian with the generalized symmetry breaking term
term $H_{\Delta}$
\begin{eqnarray}
H&=&H_0+H_{\Delta}
\label{eq:H}
,
\end{eqnarray}
where $H_0$ can be one of the Hamiltonians for correlated electrons systems.
In this paper we will restrict $H_0$ to the Hubbard model defined as
\begin{eqnarray}
H_0&=&-t\sum_{i=1}^{L-1}\sum_{\sigma} c^\dagger_{i+1,\sigma} c_{i\sigma} + h.c. 
\nonumber\\
&&+U \sum_{i=1}^L \left(n_{i\uparrow} - {1\over 2}\right)\left(n_{i\downarrow} - {1\over 2}\right)
\label{eq:H0}
\end{eqnarray}
where $c_{i\sigma}$ are fermion operators and $L$ is the system size.
The symmetry breaking term $H_{\Delta}$ is defined as
\begin{eqnarray}
H_{\Delta}&=& \sum_{i\sigma} \Delta_{i\sigma}^* c_{i\sigma}
+ \Delta_{i\sigma} c_{i\sigma}^\dagger
\end{eqnarray}
where $\Delta_{i\sigma}$ are considered as external fields at this
stage.  This model with nonzero $\Delta_{i\sigma}$ breaks the particle
number conservation because $H_{\Delta}$ is not commutable with the
total number of particles: $[H_\Delta, N_{\rm tot}]\neq 0,$ where
$N_{\rm tot} = \sum c^\dagger_{i\sigma} c_{i\sigma}$.  This comes from
the fact that $H_{\Delta}$ breaks the $U (1)$ symmetry, where the
global $U (1)$ rotation defined as $c_{i\sigma} \rightarrow
\e^{i\theta} c_{i\sigma}$.

Generally, external fields break some symmetry as magnetic fields
break spin rotational symmetry.  In addition to the $U (1)$
symmetry-breaking, $H_{\Delta}$ also breaks the $SU (2)$
spin-rotational symmetry, while the Hubbard model without magnetic
field is an $SU (2)$ invariant.  The global $SU (2)$ rotation is defined
as $\vec{c}_i \rightarrow U \vec{c} _i $, where $^t \vec{c}_i = ^t \left (
{c_{i\uparrow }}
,
{c_{i\downarrow }}
\right)$ and $\det U=1$.  
The symmetry breaking term transforms under the $SU(2)$ rotation as
\begin{eqnarray*}
H_\Delta(\vec{\Delta}_i) &=& \sum_i \vec{c}_i^\dagger \vec{\Delta}_i + \vec{\Delta}_i^\dagger \vec{c}_i
\\
\rightarrow H_\Delta' 
&=& H_\Delta(U^\dagger \vec{\Delta}_i) 
,
\end{eqnarray*}
where $^t \vec{\Delta}_i= ^t \left(
{\Delta_{i\uparrow }}
,{\Delta_{i\downarrow }}
\right) $.  The symmetry is recovered only if $\Delta_{i\sigma} =0$
for all $i, \sigma$.

\subsection{Particle-Hole Symmetry}
Let us suppose that $H_0$ satisfies the particle-hole symmetry, i.e.,
$H_0$ is invariant under the usual particle-hole
transformation on the tight-binding model: $c_{i\sigma} \rightarrow
(-1)^i c_{i\sigma}^\dagger$.  In other words, $H_0$ is commute with a
anti-unitary operator $\Theta$\cite{cond-mat.0603230}, defined as
$\Theta = K U_{{\rm ph}}$, where $K$ is a complex conjugation and
$U_{{\rm ph}}$ is the unitary operator defined as $U_{{\rm ph}} = \i^L
\prod_{i\sigma} (c_{i\sigma}+ (-1)^i c^\dagger_{i\sigma})$.  
It satisfies
\begin{eqnarray*}
\Theta^{-1} c_{i\sigma} \Theta = U_{{\rm ph}}^{-1} c_{i\sigma} U_{{\rm ph}} =
(-1)^i c_i^\dagger
,
\end{eqnarray*}
and one can show
\begin{eqnarray*}
\Theta^{-1} H_0 \Theta &=& H_0
,
\\
\Theta^{-1} H_\Delta(\vec{\Delta}_i) \Theta &=& H_\Delta\left((-1)^i \vec{\Delta}_i\right)
,
\end{eqnarray*}
where $t$ can be complex here and will be fixed $t=1$ as the unit of
energy in numerical calculations.  The symmetry-breaking term
$H_{\Delta}$ with $\vec{\Delta}_{i}=(-1)^i \vec{\Delta}_{i}$ is also
invariant under the particle-hole transformation.  Then, one can
construct a Hamiltonian $H=H_0+H_\Delta$ which preserves the
particle-hole symmetry but breaks $U (1)$ and $SU (2)$ symmetries.  It
is easy to show total number of electrons of this Hamiltonian is half
filled when the ground state is unique.  The proof is following:
because of the particle-hole symmetry $\Theta^{-1} H \Theta = H$,
$\Theta^{-1} \ket{gs}$ is also the ground state: $H \Theta^{-1}
\ket{gs}= E_{gs} \Theta^{-1} \ket{gs}$.  Since the ground state is
unique, $\ket{gs}$ is equal to $\Theta^{-1} \ket{gs}$ except for a
phase factor.  Then one gets
\begin{eqnarray}
  \label{eq:N}
 \bra{gs} N_{\rm tot} \ket{gs} &=& \bra{gs}\Theta^{-1} N_{\rm tot} \Theta \ket{gs} 
,
\end{eqnarray}
and $N_{\rm tot}$ satisfies
\begin{eqnarray}
  \label{eq:N2}
  \Theta^{-1} N_{\rm tot} \Theta &=& 2 L - N_{\rm tot}
.
\end{eqnarray}
From eqs.(\ref{eq:N}) and (\ref{eq:N2}), it is deduced that total number of
electrons is half filled, $\langle N_{\rm tot} \rangle = L$.
It might be interesting to remind the reader that the half-filled
Hamiltonian with the $U (1)$ symmetry breaking term has some analogy
to a finite double occupancy in the half-filled case discussed in the
gossamer superconducting theory proposed by
Laughlin\cite{cond-mat.0209269}.

\subsection{Hidden even-odd Parity conservation}
To handle the fermion sign especially by the DMRG which will be
discussed in the next section, let us consider a following extension
of the Hilbert space by the canonical transformation
\cite{PRB.51.1743}: $$c_{i\sigma}\rightarrow
\tilde{c}_{i\sigma}=(\alpha + \alpha^\dagger) c_{i\sigma}
,$$ where
$\alpha$ is additional annihilation operator of a spinless fermion and
satisfies
\begin{eqnarray*}
\{\alpha,c_{i\sigma}\} = 0,\;\;\;\; \{\alpha^\dagger ,c_{i\sigma}\} =0
.
\end{eqnarray*}
Canonicality of this transformation can be easily shown as
$\{\tilde{c}_{i\sigma}, \tilde{c}_{i'\sigma'}\}= - \{c_{i\sigma},
c_{i'\sigma'}\} =0$ and $\{\tilde{c}_{i\sigma},
\tilde{c}^\dagger_{i'\sigma'}\}= \{c_{i\sigma},
c^\dagger_{i'\sigma'}\} =\delta_{ii'}\delta_{\sigma\sigma'}$.
Moreover, when we denote the fermion operators in $H_0$ as $H_0
(\{c_{i\sigma}\})$, $H_0$ is invariant under this canonical
transformation:
\begin{eqnarray*}
H_0 = H_0(\{c_{i\sigma}\}) &\rightarrow& 
\widetilde{H_0}=H_0(\{\tilde{c}_{i\sigma}\}) = H_0(\{c_{i\sigma}\})
\end{eqnarray*}
because $H_0$ is made of invariant operators like
$\tilde{c}^\dagger_{i\sigma}\tilde{c}_{i'\sigma'} =
c^\dagger_{i\sigma}c_{i'\sigma'}$.
The term $\widetilde{H_\Delta}$,
however, is modified as below,
\begin{eqnarray*}
H_\Delta &\rightarrow&
\widetilde{H_\Delta}=\sum_{i\sigma} \Delta_{i\sigma}^* (\alpha + \alpha^\dagger) c_{i\sigma}
+ \Delta_{i\sigma} c^\dagger_{i\sigma}(\alpha + \alpha^\dagger) 
.
\end{eqnarray*}
This formula implies that $\alpha$ site is a ``environment bath'' site
in the spirit of the Dynamical Mean Field Theory(DMFT)\cite{RMP.68.13}.
Since single fermion operators in $H$ become bilinear,
$\widetilde{H}$ conserves a even-odd parity of the total particle
number defined below.

The operator of the total particle number $N_{\rm tot}$ is invariant
under the transformation, i.e., $\widetilde{N}_{\rm tot} = N_{\rm
  tot}$.  The total Hamiltonian $\widetilde{H}$ does not conserve
$\widetilde{N}_{\rm tot}$ nor the total particle number including
alpha $\widetilde{N}_{\alpha} = \widetilde{N}_{\rm tot} +
\alpha^\dagger\alpha$.  That is
$[\widetilde{H},\widetilde{N}_{\alpha}]\neq 0$.  We note that such
operators with $\alpha$ as $\widetilde{N}_{\alpha}$ has no
correspondent operator, for example, $N_\alpha$ which was defined
before the transformation.

The parity operator of $\widetilde{N}_{\alpha}$ is defined as
\begin{eqnarray}
\label{eq:parity}
  \widetilde{P}&=&\e^{\i\pi \widetilde{N}_{\alpha}}= \e^{\i \pi \alpha^\dagger \alpha} \prod_{i\sigma} \e^{\i \pi c^\dagger_{i\sigma} c_{i\sigma}}
  ,
\end{eqnarray}
and satisfies $\widetilde{P}^\dagger \widetilde{P} = 1$ and
$\widetilde{P}=\widetilde{P}^\dagger$.  One can show the even-odd
parity of $\widetilde{N}_{\alpha}$ is conserved, i.e.,
\begin{eqnarray*}
[\widetilde{H},\widetilde{P}]=0.  
\end{eqnarray*}

Since $\widetilde{P}$ is conserved, one can take simultaneous eigen
states of $\widetilde{H}$ and $\widetilde{P}$ as
\begin{eqnarray*}
\widetilde{H} \ket{\widetilde{\Psi}(E,p)} &=& E\ket{\widetilde{\Psi}(E,p)}
,
\\
\widetilde{P} \ket{\widetilde{\Psi}(E,p)} &=& p\ket{\widetilde{\Psi}(E,p)}
,
\end{eqnarray*}
where $E$ is eigen energy and $p$ is $\pm 1$, because
$\widetilde{P}^2$ is identity as a operator and $\widetilde{P}$ is
unitary.  The ground states $\ket{\widetilde{\Psi}(E,p)}$ for
$\widetilde{H}$ is doubly degenerated if the ground state $\ket{\Psi
  (E)}$ for $H$ is unique.  As shown in Appendix. \ref{sec:ap:ct}, one
can show that expectation value of an arbitrary operator ${\cal O}$
can be written as $\bra{\Psi (E)}{\cal O}\ket{\Psi (E)} =
\bra{\widetilde{\Psi}(E;+)}\widetilde{\cal
  O}\ket{\widetilde{\Psi}(E;+)}
=\bra{\widetilde{\Psi}(E;-)}\widetilde{\cal
  O}\ket{\widetilde{\Psi}(E;-)}$.  This means that any expectation
value for the system $H$ can be obtained in the system
$\widetilde{H}$.

\section{Method\label{sec:method}}
To study the present 1D strongly correlated system without
conservation of the total particle number, the DMRG method is used.
As we implied in the previous section, the conservation of even-odd
parity of particle number is required to calculate the fermion sign in
the DMRG algorithm.  In this section we will illustrate detail of the
DMRG algorithm and mention the fermion sign.  We note that the DMRG
method has been applied to the different model which does not conserve
total number of particles but conserves it's parity, called the BCS
pairing Hamiltonian\cite{PRL.83.172}.

First of all, we describe the iterative procedure of the DMRG for the
Hamiltonian $\widetilde{H}$ with $\alpha$ site.
\begin{figure}
\resizebox{7cm}{!}{\includegraphics{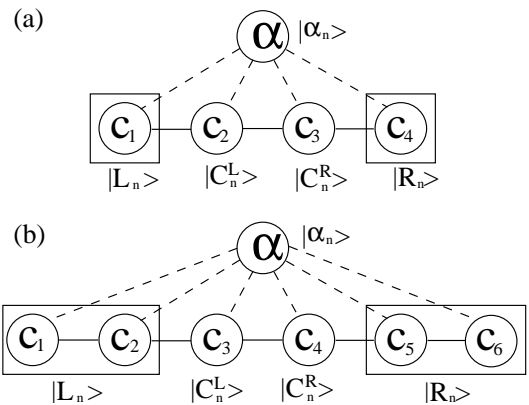}}
\caption{The system in first iterative procedure for infinite system
  algorithm of the DMRG with the bath site ``$\alpha$''.  The system
  size is enlarged from (a) $L=4$ to (b) $L=6$ .  The $\alpha$ site,
  left block, center-left site, center-right site and right block are
  represented by indices $\ket{\alpha_n}, \ket{L_n}, \ket{C^L_n},
  \ket{C^R_n}$ and $\ket{R_n}$.}
\label{fig:D}
\end{figure}
Figure \ref{fig:D} illustrates systems in first iterative procedure
with enlarging the system size from $L=4$ to $L=6$.  As seen in
fig. \ref{fig:D}(b), the hopping terms between $\alpha$ site and each
sites in $\widetilde{H}_\Delta$ become long range in the successive
elongation.  Generally speaking, long-range hopping terms such as
$c^\dagger_1 c_L$ enlarge a numerical error but the numerical error is
reduced because the $\alpha$ site is not renormalized in the iterative
procedure.

Next, to calculate the fermionic system, one should take care of the
fermion sign\cite{JPA.36.381}.  When the two local operators $\hat{A},
\hat{B}$ are represented in bases $\ket{A_n}, \ket{B_n}$, one can
usually get it's product $\bra{A_n B_m} \hat{A} \hat{B} \ket{A_{n'}
  B_{m'}}=\pm \bra{A_n} \hat{A}\ket{A_{n'}} \bra{B_m}\hat{B}
\ket{B_{m'}}$, where the coefficient $\pm$ is the fermion sign.  This
formula is valid if the states, $\ket{A_n}$ and $\ket{B_n}$, have a
fixed even-odd parity of particle number and operators $\hat{A},
\hat{B}$ conserve the even-odd parity.  Otherwise, states or operators
are modified.  For example such states as $(1+c^\dagger) \ket{0}$ may
change to $(1-c^\dagger) \ket{0}$ when the states get the fermion
sign.  This change is hard to followed in the DMRG procedure.  This is
the reason why conservation of even-odd parity is required by the
DMRG.

The canonical transformation makes it possible to calculate the
fermion sign because all operators conserve the even-odd parity as
shown in eq.(\ref{eq:POaO}) in Appendix. \ref{sec:ap:ct}.  Moreover,
local bases, $\ket{L_n}, \ket{C^L_n}, \ket{C^R_n}, \ket{R_n}$ and
$\ket{\alpha_n}$, have fixed parity as illustrated below.  

Then, all local bases in each steps should have fixed even-odd parity
of particle number.  We explain this with one step of the iterative
procedure below.
\begin{enumerate}
\item Here we suppose local bases have fixed parity of particle number
  as in $L=4$ system.
\item Make the matrix elements of the Hamiltonian
  $\widetilde{H}$ from local operators represented by local bases,
  taking care of the fermion sign.  The Hamiltonian $\widetilde{H}$ is
  block diagonalized and has even and odd parity sector.
\item Calculate the ground states
  $\ket{\widetilde{\Psi}(E_{gs},p)}$ and it's energy $E_{gs}$,
  where $p=\pm$ is even-odd parity of particle number.
\item Make the matrix elements of four density matrices $\rho_{L/R,\pm}$;
  \begin{eqnarray}
    \label{eq:rho}
\rho_{L,\pm} = 
\Tr_{R,C^R,\alpha} \rho_\pm
,\;\;\;\;
\rho_{R,\pm} = \Tr_{L,C^L,\alpha}  \rho_\pm 
,
  \end{eqnarray}
  where $\rho_\pm = \ket{\widetilde{\Psi}(E_{gs},\pm)}
  \bra{\widetilde{\Psi}(E_{gs},\pm)}$.  One can show that these
  density matrices are block diagonalized into even and odd parity
  sector, i.e., $[\rho_{L/R,\pm},\e^{\i\pi N_{L/R}}]=0$, as the Hamiltonian is.  We
  note that if the Hamiltonian conserves particle number, $[H,N_{\rm
    tot}]=0$, the density matrices for the left and right block
  conserve particle number, $[\rho_{L/R},N_{L/R}]=0$.
\item Diagonalize $\rho_{L/R,\pm}$ and select the lowest $m$
  eigenvalues and their eigenvectors called renormalized bases, which
  have fixed parity of particle number because $\rho_{L/R,\pm}$
  conserve the parity.
\item Remake matrix elements of all local operators in the
  renormalized bases.  Then, renormalized bases are the next local
  bases and satisfy the supposition in the step 1.
\end{enumerate}
As described in the procedure, since $\widetilde{H}$ conserves
even-odd parity, one can show that each local base in each steps has
fixed parity of particle number.  That is, one can calculate the
fermion sign.

Finally, we note that the number of the states of left and right block
is used up to about 60 and the truncation error is less than $10^{-4}$
in the following results.  Since we deal with general
$\Delta_{i\sigma}$ which depends on the site, the DMRG method for the
random system\cite{JPSJ.65.895} is employed.

The expectation value is deduced as
\begin{eqnarray}
\langle A\rangle &=& {1\over
  2}\sum_{p=\pm} \bra{\widetilde{\Psi}(E_{gs},p)}\tilde{A}\ket{\widetilde{\Psi}(E_{gs},p)}
,
\end{eqnarray}
in order to avoid the numerical error.  

\section{Results\label{sec:results}}
In the following, we will show results of a total-charge fluctuation and a particle
number in \S \ref{sec:totalcharge}, which are the direct effect of
existence of external fields $\Delta_{i\sigma}$.  In \S
\ref{sec:interchaincouplings} we will show tentative demonstration of
the mean-field theory to deal with the interchain hopping.

We note that we restrict the system size $L$ to even to obtain the
unique ground state $\ket{\Psi (E)}$ for $H$.  In the DMRG
calculation, $\widetilde{H}$ is used and the ground states
$\ket{\widetilde{\Psi}(E,p)}$ for $\widetilde{H}$ is doubly
degenerated.  This degeneracy is artificial due to the canonical
transformation.

\subsection{Total-charge fluctuation \label{sec:totalcharge}}
When the Hamiltonian $H_0$ is commutable with $\hat{N}$, the ground
state for $H_0$ has a fixed number of electrons and the total-charge
fluctuation, $\Delta N = \sqrt{ \langle \hat{N}^2\rangle - \langle
  \hat{N}\rangle^2}$, is zero when the ground state is unique.  On the
other hand, since external fields $\Delta_{i\sigma}$ breaks the $U
(1)$ symmetry, the total-charge fluctuation becomes finite.  The
ground state for nonzero $\Delta_{i\sigma}$ becomes a linear
combination of electron-doped states and hole-doped states.  It
implies, even at half-filling, ``effective carriers '' are introduced
by the nonzero $\Delta_{i\sigma}$.

Since large charge gap prefers no fluctuation,
when the Coulomb interaction $U$ becomes infinite, doped states are
not allowed at half-filling and $\Delta N$ becomes zero.  
\begin{figure}[h]
\resizebox{9cm}{!}{\includegraphics{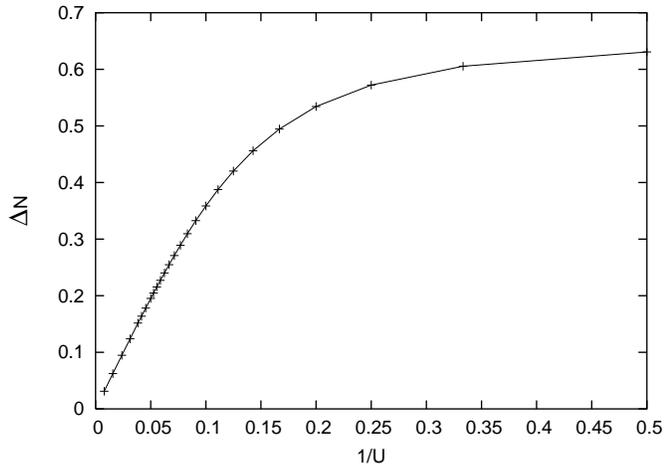}}
\caption{An example of total charge fluctuation $\Delta N$ as a
  function of $1/U$ for system-size $L=4$ at half-filling with the
  Hamiltonian $H=H_0+H_\Delta$ at $t=1$ and particle-hole symmetric
  $\Delta_{n\sigma}=\cos(n\pi/4)$, where $\langle N_{\rm tot} \rangle
  = L$ is satisfied numerically.  This figure is only for the small
  system and relatively large $\Delta_{i\sigma}$ but the qualitative
  character that $\Delta N$ is proportional to $1/U$ is general.  }
\label{fig:1}
\end{figure}
In Fig.\ref{fig:1}, total charge fluctuation $\Delta N$ as a function
of $1/U$ is plotted as an example of the particle-hole symmetric
$\Delta_{i\sigma}$.  Infinite $U$ gives no fluctuation $\Delta N=0$
and finite total-charge fluctuation is proportional to $1/U$.  This
figure is only for the small system and relatively large
$\Delta_{i\sigma}$ but the qualitative character that $\Delta N$ is
proportional to $1/U$ is general.

It should be noted that maximum of total-charge fluctuation $\Delta N$
is order 1 as fig.\ref{fig:1} shows $\Delta N \sim {\cal O}(1)$ in
large $1/U$ region. That is, the present model can not reproduce the
BCS ground state with $\Delta N \sim {\cal O}(L^{1/2})$.  This is important
for the charge compressibility defined as $\kappa (\mu) = {\partial n
  \over \partial \mu} = {1\over L}{\partial \langle N_{\rm tot}\rangle
  \over \partial \mu}$.  Let us summarize this property for the Mott
insulator and the metal: in the Mott insulating state, the charge
compressibility diverges at $\mu=\pm \Delta_c/2$ and $\kappa (\mu)=0$
at $-\Delta_c/2 <\mu< \Delta_c/2$, where $\Delta_c$ is the charge
gap\cite{PRL.20.1445}.  In the metallic case, the charge
compressibility is proportional to the density of states at the Fermi
energy.  In non-zero $\Delta_{i\sigma}$ case, the charge
compressibility at half-filling becomes finite but is expected to
become zero at $\mu=0$ in the limit $L\rightarrow \infty$ because the
magnitude of $\Delta N$ is order ${\cal O}(1)$.

\subsection{inter chain
  hopping as mean fields\label{sec:interchaincouplings}}
In this subsection, let us consider the interchain hopping of 1D Mott
insulators.  When we take an ansatz of the mean field type, external
fields $\Delta_{i\sigma}$ can be determined self-consistently as
$-t_{\perp} c^\dagger_{i\sigma} c_{i_\perp\sigma}\sim
c^\dagger_{i\sigma} \Delta_{i\sigma}$.  We adopt the self-consistent
equation:
$$\Delta_{i\sigma} = - t_{\perp} \langle c_{i\sigma} \rangle
.$$ Although the interchain hopping also gives rise to the effect of
the band structure and the dimensionality, the effect of the band
structure is not taken into account in this approach.  The meaning of
$t_{\perp}$ in the self-consistent equation is the strength of charge
fluctuation in the perpendicular direction with the general band
structure.

In the DMRG method we used the transformed Hamiltonian, $\tilde{H}$,
and the transformed self-consistent equation as $\Delta_{i\sigma} = -
t_{\perp} \langle (\alpha + \alpha^\dagger) c_{i\sigma} \rangle$.  We
note that $\Delta_{i\sigma}$ is limited to real number for simplicity.
\begin{figure}
\resizebox{9cm}{!}{\includegraphics{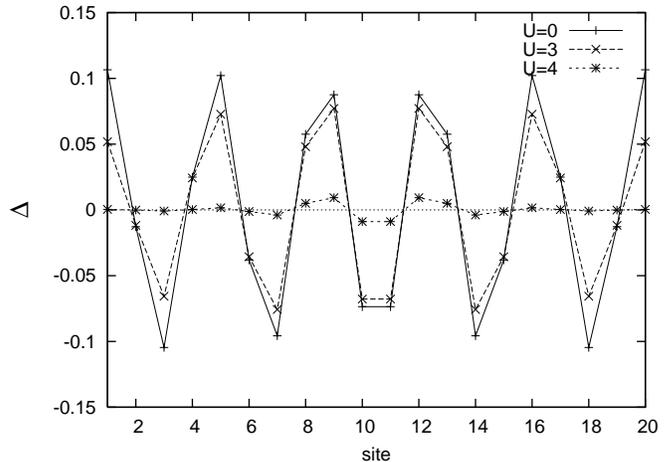}}
\caption{Converged $\Delta_{i\sigma}=\Delta_{i}$ as a real function of
  site $i$ with varying $U=0, 3, 4$. The system size is $L=20$.}
\label{fig:4}
\end{figure}
In Fig.\ref{fig:4} some result of converged $\Delta_{i\sigma}$ after
the self-consistent loop for $L=20$ are plotted.  $\Delta_{i\sigma}$
converged smaller value as $U$ increased, which implies a quantum
phase transition from nonzero $\Delta_{i\sigma}$ to zero
$\Delta_{i\sigma}$.

To clarify the transition, one can define the stabilization energy
$\Delta E=E(0)-E(\Delta_{i\sigma})$, which means the energy gain due
to charge fluctuation in the perpendicular direction, where
$E(\Delta_{i\sigma})$ is the ground state energy with
converged $\Delta_{i\sigma}$.  There are two simple limits: infinite $U$ limit
and small $t_\perp$ limit.  In both case $\Delta_{i\sigma}$ converged
to about zero and the 1D Mott insulator is realized.
\begin{figure}
\resizebox{9cm}{!}{\includegraphics{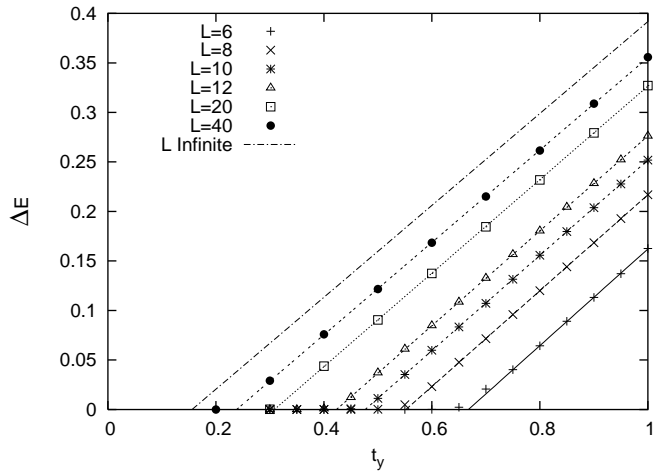}}
\caption{Stabilization energy $\Delta E=E (0)-E (\Delta_{i\sigma})$ as
  a function of $t_\perp$ with converged $\Delta_{i\sigma}$ for
  $U/t=2$.  Points are obtained by the DMRG and lines are fitted with
  a linear function.}
\label{fig:2}
\end{figure}
In Fig.\ref{fig:2} the stabilization energy is plotted as a function
of $t_\perp$.  Increasing $t_\perp$ means that length between chains
is changing more closely, which corresponds to applying pressure.  In
small $t_\perp$ region, $\Delta E$ becomes zero, which is identified
as the 1D Mott insulator phase.  Actually, converged
$\Delta_{i\sigma}$ and $\Delta N$ are zero there.  There is a
transition from the 1D Mott insulator phase to the symmetry breaking
phase as $t_\perp$ increase.  Extrapolated points in Fig.\ref{fig:2}
show that the transition point $t_\perp^c/t$ is about $0.16$ for
$U/t=2$.  Since the charge gap of Mott chain of $U/t=2$ is
$\Delta_c=0.17$\cite{PRL.20.1445}, the naive criterion $t_\perp^c \sim
\Delta_c$\cite{PRL.87.276405} is reasonable in this analysis.

\section{Conclusion\label{sec:conc}}
In conclusion, we have studied the symmetry breaking of $U (1)$ charge
and $SU (2)$ spin due to external fields $\Delta_{i\sigma}$, which
lead to nonzero total-charge fluctuation $\Delta N$.  Finite $\Delta
N$ means the coherent ground state is a linear combination of
electron-doped states and hole doped states.  We have applied the DMRG
method to the particle-hole symmetric Hubbard chain with
$\Delta_{i\sigma}$ and have actually demonstrated that the
total-charge fluctuation $\Delta N$ at zero-temperature is linear with
respect to $1/U$ even at half-filled case $\langle N_{\rm tot}\rangle
= L$.

Considering $\Delta_{i\sigma}$ as a mean field of interchain-hopping
tentatively, we have obtained the quantum phase transition from the 1D
Mott insulator to the symmetry breaking phase as $t_\perp$ increases,
i.e., pressure increases.  The transition point $t_\perp^c$ in
Fig.\ref{fig:2} is close to the charge gap $\Delta_c$.  In the
symmetry breaking phase, effectively doped carriers itinerate between
chains because of non-zero $\Delta_{i\sigma}$.  Since the difference
between two phases is whether the interchain hopping becomes relevant
or not, one may say this transition deconfinement
transition\cite{PRL.87.276405}.  However, since this approach is
limited to small $\Delta_{i\sigma}$, clarification of the symmetry
breaking phase is remained as future work.  The clarification is
interesting work because it is well known that destroying the Mott
phase by applying the pressure, i.e., increasing $t_\perp$, is typical
for high-$T_c$ cuprates.

As described in \S \ref{sec:results}, the magnitude of $\Delta N$ as a
function of the system size is constant while the BCS theory gives
$\Delta N\sim {\cal O}(L^{1/2})$.  This property may be related to the
fact that we dropped the anti-commutation relation between
$\Delta_{i\sigma}$ and fermion operators in the Hamiltonian.  That is,
expectation value $\langle c_{i\sigma} \rangle$ was a fermionic
operator before taking the average as a mean-field.  As pointed out in
ref. \cite{T.Giamarchi}, this fact gives the limitation of this ``mean
field'' approach.  We can find the another way to bosonize the mean
field and this result will be reported in future work.

As a technical outlook, we have used only the infinite method of the
DMRG in the self-consistent calculation.  The combination of finite
method and self-consistent loop will improve the cost of calculation
time, where self-consistent field is calculated at center block in
sweep of the finite method.  In this method, DMRG is combined with the
mean-field theory more closely.

\acknowledgments This work was in part supported by Grant-in-Aid for
Scientific Research (No.17540347) from JSPS and on Priority Areas
(Grant No.18043007) from MEXT.  YH was also supported in part by the
Sumitomo foundation.  Some of numerical calculations were carried out
on Altix3700BX2 at YITP in Kyoto University.
\appendix
\section{Canonical transformation \label{sec:ap:ct}}
To summarize properties of the canonical transformation,
we introduce the majorana fermions defined as
\begin{eqnarray*}
\alpha_+ &=& \alpha + \alpha^\dagger
\\
\alpha_- &=& -\i (\alpha - \alpha^\dagger)
,
\end{eqnarray*}
which are unitary and Hermite and satisfy anti-commutation relations
$\{ \alpha_+, \alpha_- \} = 0$, $\{ \alpha_\pm, c_{i\sigma} \} = 0$.
It can be proved that even-parity of $\widetilde{P}$ defined in
eq.(\ref{eq:parity}) is anti-commutable with $\alpha_-$:
$\{\widetilde{P}, \alpha_-\}=0$.  In addition, the canonical
transformation changes any operator for the system $H$, denoted by
${\cal O}$, into $\widetilde{\cal O}$ which satisfies
\begin{eqnarray}
  \label{eq:POaO}
  [\widetilde{P},
\widetilde{\cal O}]=[\alpha_-, \widetilde{\cal O}]=0
,
\end{eqnarray}
because $\widetilde{{\cal O}}$ consists of $c_{i\sigma},
c_{i\sigma}^\dagger$ and $\alpha_+$, not of $\alpha_-$ and the
operator as a polynomial function of fermion operators does not have
terms of odd degree.  Finally, since the transformation is canonical
and the new vacuum is defined as $c_{i\sigma}\ket{\widetilde{0}} =
\alpha \ket{\widetilde{0}} =0$, it can be shown that
\begin{eqnarray}
  \label{eq:OaveInv}
  \bra{0}{\cal O} \ket{0}=\bra{\widetilde{0}}\widetilde{{\cal O}} \ket{\widetilde{0}}.
\end{eqnarray}

To define the canonical transformation of the states,
we define bases explicitly as
\begin{eqnarray}
\ket{I} := \ket{\{n_i\}} &=& \prod_{i=1}^{2L} (c_i^\dagger)^{n_i} \ket{0}
.
\end{eqnarray}
After the canonical transformation,
the bases are written as
\begin{eqnarray}
\ket{\widetilde{I}} := \ket{\widetilde{\{n_i\}}} &=& \prod_{i=1}^{2L} (\alpha_+ c_i^\dagger)^{n_i} \ket{\widetilde{0}}
.
\end{eqnarray}
Since $\alpha_+^2=1$, the bases has even-parity of $\widetilde{P}$:
$\widetilde{P}\ket{\widetilde{I}} = \ket{\widetilde{I}}$, where
$\widetilde{P}$ is defined in eq.(\ref{eq:parity}).  When we define
$\ket{\widetilde{I};+}=\ket{\widetilde{I}}$ and
$\ket{\widetilde{I};-}=\alpha_- \ket{\widetilde{I}}$, one can easily
show that a set of $2\times 4^L$ bases $\ket{\widetilde{I};\pm}$ is
the ortho-normalized complete set and the bases satisfy
$\widetilde{P}\ket{\widetilde{I};\pm}=\pm\ket{\widetilde{I};\pm}$
because of $\{\widetilde{P}, \alpha_-\}=0$.

With eq.(\ref{eq:POaO}) and $\alpha_-^2=1$, one can show
${\bra{\widetilde{I};+}}\widetilde{{\cal O}}{\ket{\widetilde{I'};+}}
={\bra{\widetilde{I};-}}\widetilde{{\cal O}}{\ket{\widetilde{I'};-}}$.
With eq.(\ref{eq:OaveInv}) one can also show $\bra{I}{\cal O}\ket{I}
={\bra{\widetilde{I};+}}\widetilde{{\cal O}}{\ket{\widetilde{I'};+}}$.
Combining them, we summarize
\begin{eqnarray}
\label{eq:Osame}
\bra{I}{\cal O}\ket{I'} &=& {\bra{\widetilde{I};+}}\widetilde{{\cal O}}{\ket{\widetilde{I'};+}}
={\bra{\widetilde{I};-}}\widetilde{{\cal O}}{\ket{\widetilde{I'};-}}
.
\end{eqnarray}
This means the block diagonalized operator has the same matrix element
for even-odd sectors.  Since equation (\ref{eq:Osame}) is true when
${\cal O}$ is the Hamiltonian $H$, the eigen vectors of $H$ and
$\widetilde{H}$ can be the same:
\begin{eqnarray}
\ket{\Psi(E)} &=& \sum_I C_I(E) \ket{I}
\\
\ket{\widetilde{\Psi}(E,\pm)} &=& \sum_I C_I(E) {\ket{\widetilde{I};\pm}}
,
\end{eqnarray}
where elements $C_I(E)$ satisfy $\bra{I}H\ket{I'}C_{I'}(E)= E
C_{I}(E)$.  We note that degenerated eigen vectors
$\ket{\widetilde{\Psi}(E,\pm)}$ satisfy
$\braket{\widetilde{\Psi}(E,p)}{\widetilde{\Psi}(E',p')}=\delta_{EE'}
\delta_{pp'}$.

We conclude that the expectation value for any operator ${\cal O}$ can
be written as
\begin{eqnarray}
  \label{eq:ap-conc}
  \bra{\Psi(E)}{\cal O}\ket{\Psi(E)} &=&
  \bra{\widetilde{\Psi}(E;+)}\widetilde{\cal O}\ket{\widetilde{\Psi}(E;+)}
  \nonumber
  \\&=&
\bra{\widetilde{\Psi}(E;-)}\widetilde{\cal O}\ket{\widetilde{\Psi}(E;-)}
.
\end{eqnarray}

\end{document}